\documentclass[prl,superscriptaddress,showpacs,twocolumn]{revtex4}
\usepackage{graphicx}
\usepackage{amsmath}
\begin{document}
\author{E. Pavarini}
\affiliation{INFM and Dipartimento di Fisica ``A. Volta'', 
             Universit\`a di Pavia, Via
             Bassi 6, I-27100 Pavia, Italy}
\author{S. Biermann}
\affiliation{Centre de Physique Th\'eorique, 
             Ecole Polytechnique, 91128 Palaiseau Cedex,
             France}
\author{A. Poteryaev}
\affiliation{NSRIM, University of Nijmegen, 
             NL-6525 ED Nijmegen, The Netherlands}
\author{A. I. Lichtenstein}
\affiliation{NSRIM, University of Nijmegen, 
             NL-6525 ED Nijmegen, The Netherlands}
\author{A. Georges}
\affiliation{Centre de Physique Th\'eorique, 
             Ecole Polytechnique, 91128 Palaiseau Cedex,
             France}
\author{O.K. Andersen}
\affiliation{Max-Planck-Institut f\"{u}r Festk\"{o}rperforschung, 
             Heisenbergstrasse 1, D-70569 Stuttgart, Germany}

\title{Mott transition and suppression of orbital fluctuations in
orthorhombic 3$d^{1}$ perovskites }

\begin{abstract}
Using $t_{2g}$ Wannier-functions, a low-energy Hamiltonian is derived for
orthorhombic $3d^{1}$ transition-metal oxides. Electronic correlations are
treated with a new implementation of dynamical mean-field theory for
non-cubic systems. Good agreement with photoemission data is obtained. The
interplay of correlation effects and cation covalency (GdFeO$_{3}$-type
distortions) is found to suppress orbital fluctuations in LaTiO$_{3},$ and
even more in YTiO$_{3}$, and to favor the transition to the insulating state.
\end{abstract}

\maketitle

Transition-metal perovskites have attracted much interest because of their
unusual electronic and magnetic properties arising from narrow 3$d$ bands
and strong Coulomb correlations \cite{imada}. The 3$d^{1}$ perovskites are
particularly interesting, since seemingly similar materials have very
different electronic properties: SrVO$_{3}$ and CaVO$_{3}$ are correlated
metals with mass-enhancements of respectively 2.7 and 3.6 \cite{optics}
while LaTiO$_{3}$ and YTiO$_{3}$ are Mott insulators with gaps of
respectively 0.2 and 1\thinspace eV \cite{mottgap}.

In the Mott-Hubbard picture the metal-insulator transition occurs when the
ratio of the on-site Coulomb repulsion and the one-electron bandwidth
exceeds a critical value $U_{c}/W,$which increases with orbital degeneracy 
\cite{marcelo,Erik}. In the ABO$_{3}$ perovskites the transition-metal ions
(B) are on a nearly cubic (orthorhombic) lattice and at the centers of
corner-sharing O$_{6}$ octahedra. The 3$d$ band splits into
$pd\pi $-coupled $t_{2g}$ bands and 
$pd\sigma $-coupled $e_{g}$\thinspace bands, of which the former lie lower,
have less O character, and couple less to the octahedra than the latter.
Simplest theories for the $d^{1}$ perovskites\thinspace \cite{imada}\ are
therefore based on a Hubbard model with 3 \emph{degenerate}, $\frac{1}{6}$
-filled $t_{2g}$\thinspace bands per B-ion, and the variation of the
electronic properties along the series is ascribed to a progressive
reduction of $W$ due to the increased bending of the $pd\pi $ hopping paths
(BOB bonds).

This may not be the full explanation of the Mott transition however, because
a splitting of the $t_{2g}\,$levels can effectively lower the degeneracy. In
the correlated metal, the relevant energy scale is the reduced bandwith
associated with quasiparticle excitations. Close to the transition, this
scale is of order $\sim ZW$, with $Z\sim 1-U/U_{c}\,$, and hence much
smaller than the original bandwith $W$. A level splitting by merely $ZW$ is
sufficient to lower the effective degeneracy all the way from three-fold to
a non-degenerate single band\cite{Manini}. 
This makes the insulating state more favorable by reducing 
$U_{c}/W$\cite{Manini,Erik}.
Unlike in $e_{g}
$-band perovskites, such as LaMnO$_{3},$ where large (10\%) cooperative
Jahn-Teller (JT) distortions of the octahedra indicate that the orbitals are
spatially ordered, in the $t_{2g}$-band perovskites the octahedra are 
almost perfect. The $t_{2g}$\thinspace orbitals have therefore often been
assumed to be degenerate. If that is true, it is conceivable that quantum
fluctuations lead to an orbital \emph{liquid}~\cite{Khal} rather than
orbital ordering. An important experimental constraint on the nature of the
orbital physics is the observation of an isotropic, small-gap spin-wave
spectrum in both insulators\thinspace \cite{Keim}. This is remarkable
because LaTiO$_{3}$ is a G-type antiferromagnet with $T_{N}$=140\thinspace K, 
$m$=0.45$\,\mu _{B},$ and a 3\% JT stretching along 
$\mathbf{a}$ \cite{lastr2}, while YTiO$_{3}$ is a ferromagnet with 
$T_{C}$=30\thinspace K, 
$m_{0}\sim $0.8$\mu _{B},$ and a 3\% stretching along 
$\mathbf{y}$ on sites 1 and 3, and $\mathbf{x}$ on 2 and 4 \cite{ystr} (see
Fig.\thinspace 1).

We shall find that the $t_{2g}$ degeneracy \emph{is lifted} at the classical
level. This is not due to the small JT distortions via OB $pd\pi $-coupling,
but to the GdFeO$_{3}$-type distortion which tilts the corner-sharing
octahedra around the \textit{b}-axis 
(by 0,\thinspace 9,\thinspace 12, and 20$^{\circ }$) and
rotates them around the \textit{c}-axis (by 0,\thinspace 7,\thinspace 9, and 13$^{\circ }$), as
we progress from cubic SrVO$_{3}$ via CaVO$_{3}$ and LaTiO$_{3}$ to YTiO$_{3}
$ \cite{srstr,castr,lastr2,ystr}. This distortion is driven by the increasing
oxygen-cation (OA) $pd\sigma $-covalency, and it primarily pulls closer
4 of the 12 oxygens neighboring a given cation\cite{footOA}. Moreover, 2
to 4 of the 8 cations neighboring a given 
B ion are pulled
closer~\cite{footAB}. The $t_{2g}$ orbitals couple to the OA distortion via
oxygen (BOA$\,dp\pi $-$pd\sigma $), and they couple directly (AB $dd\sigma )$
to the AB distortion. As seen in Fig.\thinspace 1, the orthorhombic GdFeO$
_{3}$-type distortion also leads to quadrupling of the cell. 
These findings are consistent with conclusions drawn in the most recent
model Hartree-Fock study\cite{recent}.  The correct magnetic orders in
LaTiO$_{3}$ and YTiO$_{3}$ were also obtained with the 
LDA+\textit{U} method~\cite{Terakura}. 
The predicted
moment and orbital order in YTiO$_{3}$ were confirmed by NMR \cite{Itho} and
neutron scattering \cite{neutrons}, but not in LaTiO$_3$.
These static mean-field methods are not appropriate for the metallic
systems, however.

\begin{figure}[th]
\label{orbitals} \vspace*{-.1cm}
\par
\begin{center}
\rotatebox{270}{\includegraphics[width=0.43\textwidth]{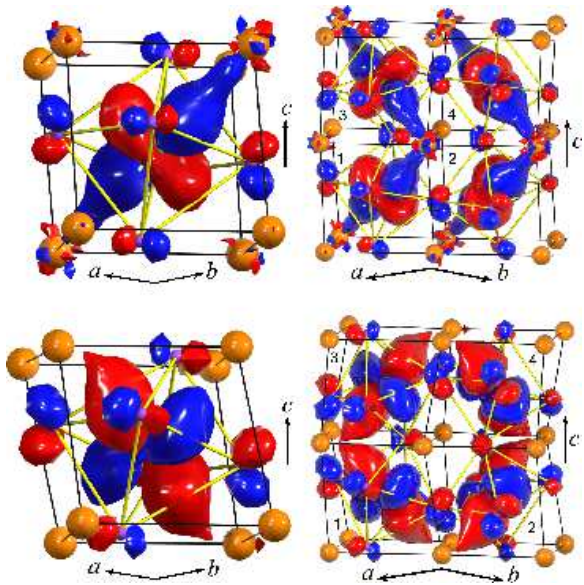}}
\end{center}
\par
\vspace*{-.35cm}
\caption{\textit{Pbnm} primitive cell (right), subcell\thinspace1 (left),
and the occupied $t_{2g}$ orbitals for LaTiO$_{3}$ (top) and YTiO$_{3}$
(bottom) according to the LDA+DMFT calculation. The oxygens are violet, the
octahedra yellow and the cations orange. In the global, cubic $xyz$-system
directed approximately along the Ti-O bonds, the orthorhombic translations
are $\mathbf{a}\mathrm{=}\left( 1,-1,0\right) \left( 1+\protect\alpha\right)
,\ \mathbf{b}\mathrm{=}\left( 1,1,0\right) \left( 1+\protect\beta\right) ,$
and $\mathbf{c}\mathrm{=}\left( 0,0,2\right) \left( 1+\protect\gamma\right)
, $ with $\protect\alpha,\protect\beta,\protect\gamma$ small. The Ti sites 1
to 4 are: \textbf{a}/2, \textbf{b}/2, (\textbf{a+c})/2, and (\textbf{b+c)}%
/2. The La(Y) $ab$-plane is a mirror $\left( z\leftrightarrow-z\right) $,
and so is the Ti $bc$-plane $\left( x\leftrightarrow y\right) $ when
combined with the translation (\textbf{b-a})/2.
See: http://www.mpi-stuttgart.mpg.de/andersen/cm/0309102.html
}
\end{figure}

In this letter, we shall (i) present a new implementation of a many-body
method~\cite{DMFT,ldadmft}, which allows for a quantitative,
material-specific description of \textit{both} the Mott transition and the
orbital physics, and (ii) use it to explain why some of the $d^{1}$
perovskites are metallic and others are insulators, why the metals have
different mass enhancements and the insulators different gaps. Such
properties can be described by a low-energy, multi-band Hubbard Hamiltonian,%
\begin{align}
H& =H^{LDA}+\frac{1}{2}\sum\nolimits_{imm^{\prime }\sigma }U_{mm^{\prime
}}n_{im\sigma }n_{im^{\prime }-\sigma }  \notag \\
& +\frac{1}{2}\sum\nolimits_{im\left( \neq m^{\prime }\right) \sigma
}(U_{mm^{\prime }}-J_{mm^{\prime }})n_{im\sigma }n_{im^{\prime }\sigma },
\label{H}
\end{align}%
where $n_{im\sigma }=a_{im\sigma }^{+}a_{im\sigma },$ and $a_{im\sigma }^{+}$
creates an electron with spin $\sigma $ in a localized orbital $m$ at site $%
i.$ This Hamiltonian depends on how the $im\sigma $-orbitals are chosen. $%
H^{LDA}$ is the one-electron part given by density-functional theory (LDA),
which should provide the proper material dependence. Recently it has become
feasible to solve (\ref{H}) using the \emph{dynamical} mean-field
approximation (DMFT)\thinspace \cite{DMFT} and to obtain realistic physical
properties\thinspace \cite{ldadmft}. In the original LDA+DMFT
implementations it was assumed that the on-site block(s) of the
single-particle Green function is diagonal in the space of the correlated
orbitals, and these were taken as orthonormal LMTOs \emph{approximated} by
truncated and renormalized partial waves. 
\begin{figure}[th]
\label{Ldados}
\par
\begin{center}
\includegraphics[width=0.45\textwidth]{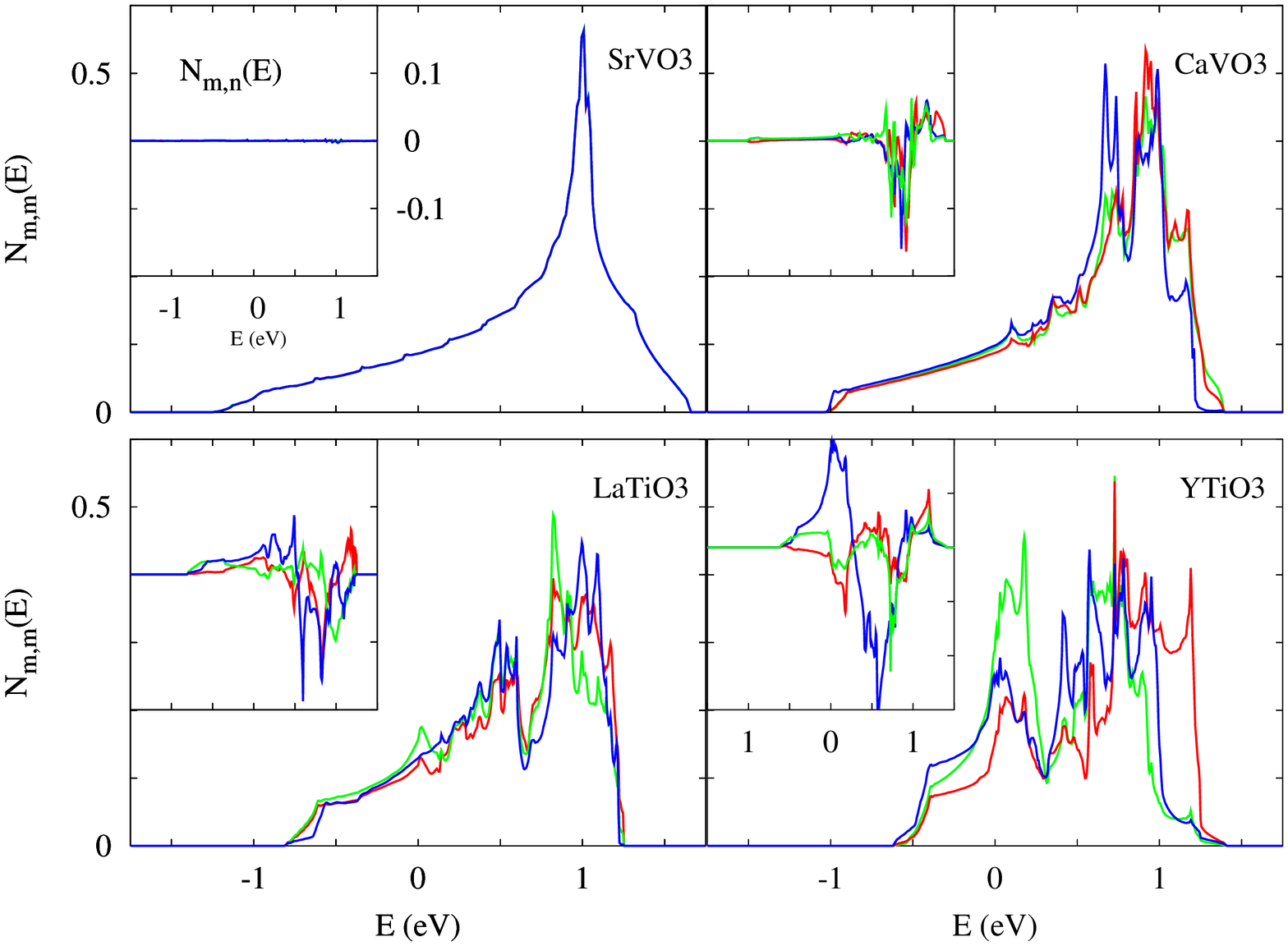}
\end{center}
\caption{$t_{2g}$ LDA DOS matrix (states/eV/spin/band) in the Wannier
representation. On-site-1 elements: $N_{xz,xz}$ (red) $N_{yz,yz}$ (green),
and $N_{xy,xy}$ (blue). Insets: $N_{yz,xz}$ (red), $N_{xz,xy}$ (green), and $%
N_{xy,yz}$ (blue). $\protect\varepsilon _{F}\equiv 0.$}
\end{figure}
Although these approximations are good for cubic $t_{2g}$ systems such as
SrVO$_{3}$\cite{Held}, they deteriorate with the degree of distortion. Our
new implementation of LDA+DMFT uses a set of localized Wannier functions in
order to construct 
a realistic Hamiltonian (\ref{H}), which is then solved by DMFT, including
the non-diagonal part of the on-site self-energy.

For an isolated set of bands, a set of Wannier functions constitutes a \emph{%
complete,} orthonormal set of orbitals with \emph{one} orbital per band. For
the $d^{1}$ perovskites we take the correlated orbitals to be three
localized $t_{2g}$ Wannier-orbitals, and in $H^{LDA}$ we neglect the degrees
of freedom from all other bands. In order to be complete, such a Wannier
orbital must have a tail with e.g.\thinspace O$\,p\pi $ and A$\,d$
characters. Our Wannier orbitals are symmetrically orthonormalized Nth-order
muffin-tin orbitals (NMTOs) \cite{nmto}, which have all partial waves other
than B$\,xy,\,yz,$ and\thinspace $zx$ downfolded. Such a $t_{2g}$ NMTO 
can have on-site $e_{g}$\thinspace character, and that
allows the orbital to orient itself after the surroundings, although $xy,$%
\thinspace $yz,$ and $zx$ refers to the global cubic axes defined in
Fig.\thinspace 1. Fourier transformation of the orthonormalized 12$\times 12$
NMTO Hamiltonian, $H^{LDA}\left( \mathbf{k}\right) ,$ yields on-site blocks
and hopping integrals. For the on-site Coulomb interactions in Eq.\thinspace
(1), we use the common assumption that, as in the isotropic case, they can
be expressed in terms of two parameters: $U_{mm}\mathrm{=}U,\ U_{mm^{\prime
}}\mathrm{=}U-2J,\ $and $J_{mm^{\prime }\left( \neq m\right) }\mathrm{=}J$ %
\cite{fresard}. From Ref.\thinspace \onlinecite{MF96}, $J$=0.68\thinspace eV
for the vanadates and 0.64\thinspace eV for the titanates. Since our
Hamiltonian involves only correlated\ orbitals, so that the number of
correlated electrons is fixed, the double-counting correction amounts to an
irrelevant shift of the chemical potential. $H$ is now solved within DMFT,
i.e.\thinspace under the assumption that the components of the self-energy
between different sites can be neglected. As a result, the self-energy can
be obtained from the solution of an effective local impurity model which
involves only 3 correlated orbitals. In contrast to all previous studies, we
take all components of the self-energy matrix $\Sigma _{mm^{\prime }}$
between different Wannier functions on a given B-site into account \cite%
{sigmaoff}. From this 3$\times $3 matrix, by use of the \textit{Pbnm}
symmetry (Fig.\thinspace 1), we construct a 12$\times $12 block-diagonal
self-energy matrix. The latter is then used together with $H^{LDA}\left( 
\mathbf{k}\right) $ to obtain the Green function at a given $\mathbf{k}$%
-point. Fourier transformation over the entire Brillouin zone yields the
local Green function associated with a primitive cell and its 3$\times 3$
on-site block is used in the DMFT self-consistency condition in the usual
manner. The 3-orbital impurity problem is solved by a numerically exact
quantum Monte Carlo scheme\thinspace \cite{hirsch}. To access temperatures
down to 770\thinspace K, we use up to 100 slices in imaginary time. 10$^{6}$
QMC sweeps and 15-20 DMFT iterations suffice to reach convergence. Finally,
the spectral function is obtained using the maximum entropy method~\cite%
{jarrell}.

We now present 
the LDA results for the four perovskites. Fig.~2 displays the on-site DOS
matrix $N_{mm^{\prime }}(\varepsilon )$ in the representation of the $xy,yz,$
and $zx$ Wannier functions. SrVO$_{3}$ is cubic and its $t_{2g}$ band with a
width $W$=2.8~eV consists of 3 non-interacting subbands, each of which is
nearly 2D and gives rise to a nearly logarithmic DOS peak. In CaVO$_{3}$, $W$
is reduced to 2.4\thinspace eV because the Wannier orbitals are misaligned
by the GdFeO$_{3}$-type distortion and because some of their O$\,2p$
character is stolen by the increased %
OA covalency, which drives this distortion. 
The energy of the $xy$%
\thinspace Wannier orbital (the center of gravity of $N_{xy,xy})$ is
80\thinspace meV lower than that of the degenerate $xz$ and $yz$ orbitals,
and small off-diagonal DOS elements appear. Going from the vanadates to the
titanates, the effects of %
OA and (A$\,d)$(B$\,t_{2g})$ covalency increase dramatically, because now A
and B are 1st rather than 3rd-nearest neighbors in the periodic table. As
consequences, the increased misalignment and loss of oxygen character
reduces the bandwidths to 2.1 and 2.0$\,$eV in LaTiO$_{3}$ and YTiO$_{3},$
and weak hybridization with the A$\,d$ bands deforms the $t_{2g}$
band.
A pseudo-gap which starts out as a splitting of
the van Hove peak in CaVO$_{3}$, deepens and moves to lower occupancy as we
progress to LaTiO$_{3}$ and YTiO$_{3}.$ The deep pseudogap in YTiO$_{3}$ is
mainly caused by the hybridization with the Y\thinspace $xy$ orbital$.$ The $%
xy,yz,\,$and$\,zx$ Wannier orbitals are now strongly mixed and
diagonalization of the on-site blocks of $H^{LDA}$ yields three
singly-degenerate levels with the middle (highest) being 140 (200)\thinspace
\begin{figure}[th]
\label{dmft}
\par
\begin{center}
\rotatebox{270}{\includegraphics[width=0.32%
\textwidth]{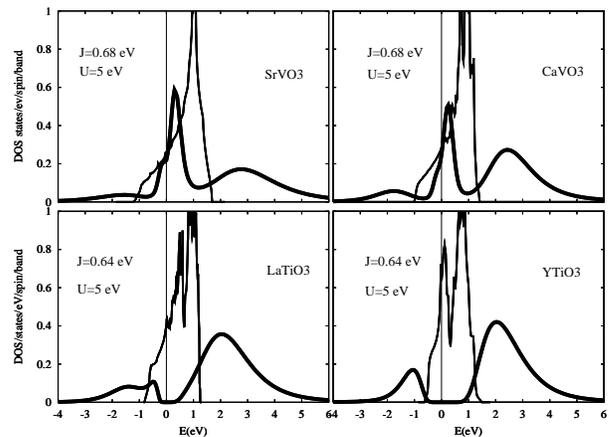}}
\end{center}
\caption{DMFT spectral function at $T=770K$ (thick line) and LDA DOS (thin
line). $\protect\mu \equiv 0.$}
\end{figure}
meV above the lowest in LaTiO$_{3},$ and 200 (330) meV in YTiO$_{3}.$ This
splitting is not only an order of magnitude smaller than the $t_{2g}$
bandwidth, but also smaller than the subband-widths, in particular for LaTiO$%
_{3}$. As a consequence, the eigenfunction for the lowest level is occupied
by merely 0.45\thinspace electron in LaTiO$_{3}$ and 0.50 in YTiO$_{3},$
while the remaining 0.55 (0.50)\thinspace electron occupies the two other
eigenfunctions. The eigenfunction on site 1 with the lowest energy is $%
0.604|xy\rangle +0.353|xz\rangle +0.714|yz\rangle $ in LaTiO$_{3}$ and $%
0.619|xy\rangle -0.073|xz\rangle +0.782|yz\rangle $ in YTiO$_{3}$. The
splittings are large compared with the spin-orbit splitting $\left( \text{%
20\thinspace meV}\right) $ and $kT,$ and they are not caused by the JT
distortions, as we have verified by turning %
them off in the calculations. %

Next, we turn to the LDA+DMFT results. Calculations were performed for
several values of $U$ between 3 and 6\thinspace eV. We found that the
critical ratio $U_{c}/W$ decreases when going along the
series: SrVO$_{3}$, CaVO$_{3}$, LaTiO$_{3}$, and YTiO$_{3}$. This is
consistent with the increasing splitting of the $t_{2g}$\thinspace levels
and indicates that the Mott transition in the $d^{1}$ series is driven as
much by the decrease of effective degeneracy as by the reduction of
bandwidth. The main features of the photoemission spectra for all four
materials, as well as the correct values of the Mott-Hubbard gap for the
insulators \cite{mottgap}, are reproduced by taking $U$ constant $\sim $5$\,$%
eV. This is satisfying, because $U$ is expected to be similar for vanadates
and titanates, although slightly smaller for the latter \cite{MF96}. In
Fig.\thinspace 3 we show the DMFT spectral functions together with the LDA
total DOS. For cubic SrVO$_{3}$ we reproduce the results of previous
calculations \cite{Liebsch,Held}: the lower Hubbard band (LHB) is around $%
-1.8$\thinspace eV and the upper Hubbard band (UHB) around 3\thinspace eV.
Going to CaVO$_{3}$, the quasiparticle peak looses weight to the LHB, which
remains at $-1.8$\thinspace eV, while the UHB moves down to 2.5\thinspace
eV. These results are in good agreement with photoemission data\thinspace %
\cite{photoemission} and show that SrVO$_{3}$ and CaVO$_{3}$ are rather
similar, with the latter slightly more correlated. Similar conclusions were
drawn in Ref.\thinspace \onlinecite{Held}. From the linear regime of the
self-energy at small Matsubara frequencies we estimate the quasi-particle
weight to be $Z$=0.45 for SrVO$_{3}$ and 0.29 for CaVO$_{3}$. For a \textbf{k%
}-independent self-energy, as assumed in DMFT, this yields $m^{\ast
}/m=1/Z=2.2$ for SrVO$_{3}$ and $3.5$ for CaVO$_{3}$, in reasonable
agreement with the optical-conductivity values $2.7$ and $3.6$\thinspace %
\cite{optics}.

For LaTiO$_{3}$ and YTiO$_{3}$ the LHB is around $-1.5$\thinspace eV, in 
accord with photoemission\thinspace \cite{photoemission2}. Despite very
similar bandwidths, the gaps are very different, 0.3 and 1\thinspace eV, and
this also agrees with experiments\thinspace \cite{mottgap}. This is
consistent with our findings that the $t_{2g}$-level splittings are smaller
and $\left( U-2J\right) _{c}/W$ is larger in LaTiO$_{3}$ than in YTiO$_{3},$
where the orbital degeneracy is effectively 1. Diagonalization of the matrix
of occupation numbers reveals that for the titanates \emph{one orbital per
site is nearly full}, in contrast to LDA. It contains 0.88\thinspace
electron in LaTiO$_{3}$ and 0.96 in YTiO$_{3}.$ The orbital polarization
increases around the metal-insulator transition and becomes complete
thereafter. Thus, for the vanadates, each orbital is approximately 1/3
occupied for all $U$ in the range 0\negthinspace to 6\thinspace eV. The 
\emph{nearly complete} orbital polarization found by LDA+DMFT for the two
insulators indicates that correlation effects in the paramagnetic Mott
insulating state considerably decrease orbital fluctuations, and makes it
unlikely that YTiO$_{3}$ is a realization of an orbital liquid\thinspace %
\cite{Khal}. In LaTiO$_{3}$ some orbital fluctuations are still active,
although quite weak. The occupied orbital in LDA+DMFT is $0.586|xy\rangle
+0.275|xz\rangle +0.762|yz\rangle $ for LaTiO$_{3}$ and $0.622|xy\rangle
-0.029|xz\rangle +0.782|yz\rangle $ for YTiO$_{3}.$ Hence, it is nearly
identical with the ones we obtained from the LDA as having the lowest
energy. For YTiO$_{3}$ our orbital is similar to the one obtained in
Ref.\thinspace \cite{Terakura} and for LaTiO$_{3}$ it is similar to the one
obtained in Refs.\thinspace \cite{lastr2,recent}. Our accurate Wannier
functions show \emph{why} these orbitals (Fig.\thinspace 1, left) have the
lowest energy: The positive (blue) lobes have bonding $3z_{111}^{2}-1=\left(
|xy\rangle +|xz\rangle +|yz\rangle \right) /\sqrt{3}$ character on the
nearest cations --those along [111]-- and the negative (red) lobes have
bonding $xy$ character on the next-nearest cations --those along [1-11]--
whose oxygen surrounding is favorable for this type of bond, i.e.\thinspace
where (O\thinspace $p)$(Y$\,xy)$ $pd\sigma $-hybridization is strong. The
former type of AB covalency dominates in LaTiO$_{3},$ while the latter
dominates in YTiO$_{3},$ where the shortest YO bond is merely 10\% longer
than the TiO bond. The difference seen (Fig.\thinspace 1, right side)
between the orbital orders in the two compounds is therefore quantitative
rather than qualitative; it merely reflects the extent to which the orbital
has the $bc$-plane as mirror. The two different JT distortions of the oxygen
square is a reaction to, rather than the cause of the difference in the
orbital orders. This difference is reflected in the hopping integrals
between nearest neighbors: $t_{x}$=$t_{y}$=99 (38) meV and $t_{z}$=105 (48)
meV for LaTiO$_{3}$ (YTiO$_{3}).$ These hoppings are fairly isotropic and
twice larger in LaTiO$_{3}$ than in YTiO$_{3}.$ Moreover, the hoppings to
the two excited orbitals are stronger in YTiO$_{3}$ than in LaTiO$_{3}.$ All
of this is consistent with LaTiO$_{3}$ being G-type anti- and YTiO$_{3}$
ferromagnetic at low temperature, and it warrants detailed future
calculations of the spin-wave spectra.

In conclusion, we have extended the LDA+DMFT approach to the non-cubic case
using ab-initio downfolding in order to obtain a low-energy Wannier
Hamiltonian. Applying this method to the Mott transition in 3$d^{1}$
perovskites, we have explained the photoemission spectra and the values of
the Mott gap without adjustable parameters, except a single value of $U.$
The Mott transition is driven by correlation effects and GdFeO$_{3}$-type
distortion through reduction, not only of bandwidth, but also of effective
orbital degeneracy. Correlation effects and cation covalency suppress
orbital fluctuations in the high-temperature paramagnetic insulating phase
of LaTiO$_{3}$ and YTiO$_{3}.$

We thank J.\thinspace Nuss,~G.\thinspace Khaliullin,~E.\thinspace Koch,~J.\thinspace Merino, M.\thinspace Rozenberg,\thinspace A.\thinspace Bringer,\thinspace
M.\thinspace Imada for~useful~discussions and the KITP Santa Barbara for hospitality
and support (NSF Grant PHY99-07949). Calculations were performed at MPI-FKF
Stuttgart and IDRIS Orsay (project No.\thinspace021393). 
S.B. acknowledges support from the CNRS and the EU 
(Contract No.\thinspace HPMF-CT-2000-00658).

\end{document}